%% file: paper.tex
\documentclass[twoside,english,aps,prb,twocolumn,showpacs,superscriptaddress]{revtex4-2}
\usepackage[utf8]{inputenc}
\usepackage{amsmath}
\usepackage{amssymb}
\usepackage{graphicx}
\usepackage{esint}
\usepackage{xcolor}
\usepackage{stmaryrd}

\makeatletter

\usepackage{amsfonts}

\makeatother

\usepackage{babel}
\begin{document}

\title{Fractional positional jumps in stochastic systems with tilted periodic double-well potentials}

\author{Martin \v{Z}onda }
\email{martin.zonda@matfyz.cuni.cz}

\affiliation{Department of Condensed Matter Physics, Faculty of Mathematics anplaced Charles University, Ke Karlovu 5, 121 16 Praha 2, Czech Republic}

\author{Wolfgang Belzig}

\affiliation{Fachbereich Physik, Universität Konstanz, D-78457 Konstanz, Germany}

\author{Edward Goldobin}

\affiliation{Physikalisches Institut, Center for Quantum Science (CQ) and LISA$^+$, 
Universitat T\"ubingen, Auf der Morgenstelle 14, D-72076 T\"ubingen,
Germany}

\author{Tom\'{a}\v{s} Novotn\'y}
\email{tomas.novotny@matfyz.cuni.cz}

\affiliation{Department of Condensed Matter Physics, Faculty of Mathematics and
Physics, Charles University, Ke Karlovu 5, 121 16 Praha 2, Czech Republic}

\date{\today}
\begin{abstract}
We present a theoretical investigation of the stochastic dynamics of a damped particle in a tilted periodic potential with a double well per period. By applying the matrix continued fraction technique to the Fokker-Planck equation in conjunction with the full counting statistics and master equation approaches, we determine the rates of specific processes contributing to the system's overall dynamics. At low temperatures, the system can exhibit one running state and two distinct locked metastable states. We focus primarily on two aspects: the dynamics of positional jumps, which are rare thermally induced particle jumps over potential maxima, and their impact on the overall velocity noise; and the retrapping process, involving the transition from the running to the locked metastable states.
We demonstrate the existence of fractional (in units of $2\pi$) positional slips that differ qualitatively from conventional $2\pi$ jumps observed in single-well systems. Fractional positional slips significantly influence the system dynamics even in regimes dominated by dichotomous-like switching between running and locked states. Furthermore, we introduce a simple master equation approach that proves effective in analyzing various stages of the retrapping process. Interestingly, our analysis shows that even for a system featuring a well-developed double-well periodic potential, there exists a broad parameter range where the stochastic dynamics can be accurately described by an effective single-well periodic model. The techniques introduced here allow for valuable insights into the complex behavior of the system, offering avenues for understanding and controlling its steady-state and transient dynamics, which go beyond or can be complementary to direct stochastic simulations.
\end{abstract}

\maketitle

\section{Introduction}

The stochastic dynamic motion of a particle in tilted periodic potentials
plays an important role in the studies of various physical phenomena
including Josephson junctions (JJs)~\cite{Likharev86,Golubov,Longobardi11,Menditto2018Evidence,ryabov2022phase}, microparticles confined in shaped laser beams~\cite{Evstigneev08,McCannNAT99,TatarkovaPRL03,CizmarPRB06,SilerNJP08,SilerNJP10,siler2018diffusing,Bellando2022giant},
dynamics of charge density waves~\cite{Gruner88}, crystal surface
melting~\cite{Frenken85,Pluis87}, ratchet and molecular motors~\cite{Julicher1997modeling,Menditto16_PRE,Hayashi2015giant}, cold atoms in optical lattices~\cite{Denisov2014Tunable}, and different biophysical
processes~\cite{Nixon96} as well as in investigations of phenomena such as anomalous diffusion and memory effects~\cite{Spiechowicz2016Transient,Goychuk2019fractional,Spiechowicz2020Diffusion,Bialas2020Colossal,Goychuk2021Nonequilibrium}. A well-paid effort has already been invested in the theoretical analysis of such systems~\cite{Risken,Elachi76,Hanggi90,LEbook}, yet there are still plenty of open questions often inspired by the recent experimental realizations in different systems, e.g.,  Refs.~\cite{Menditto2018Evidence,Bellando2022giant,Trahms2023current,Kolzer2023suppercurent}. Moreover, experimental progress in different fields called
for readdressing some old problems such as escape and retrapping
of the Brownian particle from or to potential minima~\cite{Zonda12,Germent2014,Zonda15,Cheng2015,Wang17,Spiechowicz2020Diffusion}, the statistics of thermally activated jumps of the particle by
integer multiples of $2\pi$~\cite{Little67,Melnikov91,Costantini1999treshold,Zonda15} and multistability~\cite{Spiechowicz2020Diffusion,Spiechowicz2021Arcsine,Spiechowicz2022Velocity}. 
 
In this respect, of special
interest are systems where the stochastic dynamics of the particle
is affected by a biharmonic potential containing two local minima per period
\begin{eqnarray}
U(\varphi) & = & -\varphi\,i_{b}+U_{0}(\varphi),\label{eq:Potential1}\\
U_{0}(\varphi) & = & -\alpha\cos\varphi+\frac{1}{2}\cos2\varphi,\nonumber 
\end{eqnarray}
where $i_{b}$ is a static external bias force, i.e., a potential tilt, $U_{0}(\varphi)$
is the untilted periodic double-well potential, where the coefficient $\alpha$
tunes the ratio of the first to the second harmonic contribution and
$\varphi$ is the dimensionless position. 
The potential~\eqref{eq:Potential1}
plays an important role in the theoretical description of numerous physical systems, including JJs with substantial second harmonic in the current
phase relation (CPR)~\cite{Goldobin07,Mints98,Mints00,Mints01,Goldobin11,Sickinger12,Heim13,Goldobin13,Feinberg14,Ouaussou16,Jack17,Bakurskiy17,Menditto2018Evidence,Kadlecova2019Practical,zhang2022large}
in particular in Josephson diodes~\cite{Pal2022cooper,Souto2022josephson,Trahms2023current,fomirov2022asymmetric,seleznev2024influence}
(where the variable $\varphi$ is the superconducting phase difference, i.e., the Josephson phase),
or various ratchet systems~\cite{Bartussek94,Hutchings04,Hanggi09}, molecular motors~\cite{Reimann2002,Kassem2017Artificial} and multiband superconductors~\cite{yerin2021phase}. 

A careful analysis of the
motion of the particle in the tilted double-well potential has already
led to some important results. 
For example, it explained the existence of two critical escape currents from the superconducting
to the resistive state observed experimentally for the JJs with doubly
degenerate ground state (i.e., $\varphi$ junctions)~\cite{Sickinger12}, pointed to rather nontrivial retrapping
dependencies of the Josephson phase which can lead
to a butterfly effect~\cite{Goldobin13}
and showed the existence of chaotic Josephson phase trajectories in various generalizations
of the famous RCSJ model~\cite{AlKhawaja2008,Canturk2012}. The case
$\alpha=0$, where only the second harmonic is present in the potential,
plays an important role in the studies of unconventional junctions
which can undergo the so-called $0$-$\pi$ transition~\cite{Baselmans99,Bakurskiy17}. 
Moreover, a Brownian
particle moving in the potential~\eqref{eq:Potential1} with additional harmonic terms, characterized by asymmetric mobility considering the bias force, is an archetypal model of the ratchet and diode systems~\cite{Bartussek94,Hanggi09,Pal2022cooper,Souto2022josephson}.

A crucial component in all of the above phenomena is noise. In combination with different dampings, a system with potential~\eqref{eq:Potential1}
can show a wide variety of regimes, each important for different physical
realizations. 
Here, we provide a systematic analysis of a general case. We start with the simple strong-damping parameter regime and proceed to the more complicated intermediate- and weak-damping regimes. 
We use velocity noise to identify three main dynamic
regimes~\cite{Zonda15}: the thermal-noise regime; the positional-jump (PJ) regime, and the switching regime. In each, we focus on the analysis of the dominant dynamics. In particular, we investigate the statistics of the positional jumps and their contribution to the overall velocity noise in the PJ regime and the escape and retrapping processes in the switching one. 
For this purpose, we combine the matrix continued-fraction (MCF) method with other techniques. In particular, in Sec.~\ref{sec:Methods} we introduce a combination of the MCF technique~\cite{Risken} with full counting statistics (FCS)~\cite{Ferrando92,Ferrando93,Golubev10} for the analysis of the multiple PJs.
Its biggest advantage over stochastic simulations is the straightforward access to the steady state. 
As such, it is suitable for the calculation of rates for rare events. Therefore, in the relevant regime, this method allows decomposition of the particle dynamics into independent elementary processes~\cite{Vanevic:PRL07,Vanevic:PRB08,Padurariu:PRB12}, constituted by
single or multiple positional jumps. Using this method, we demonstrate the existence of the so-called fractional PJs in
Sec.~\ref{subsec:Phase-jumps}.   
In Sec.~\ref{subsec:Switching-processes}
we investigate retrapping processes in the switching regime. Here we combine the MCF with an effective master equation approach describing the transition of the system between its three metastable regimes. On top of the steady-state studies, we also discuss in Sec.~\ref{subsec:Retrapping} a dynamical retrapping scenario. 

There is a particular conclusion of our research that is worth foreshadowing here. Namely, for a broad range of parameters, even a system with a well-developed double-well potential can be faithfully described by a single-well model.

\section{Model and Methods}
\label{sec:Methods}
\begin{figure}[ht]
	\includegraphics[width=1\columnwidth]{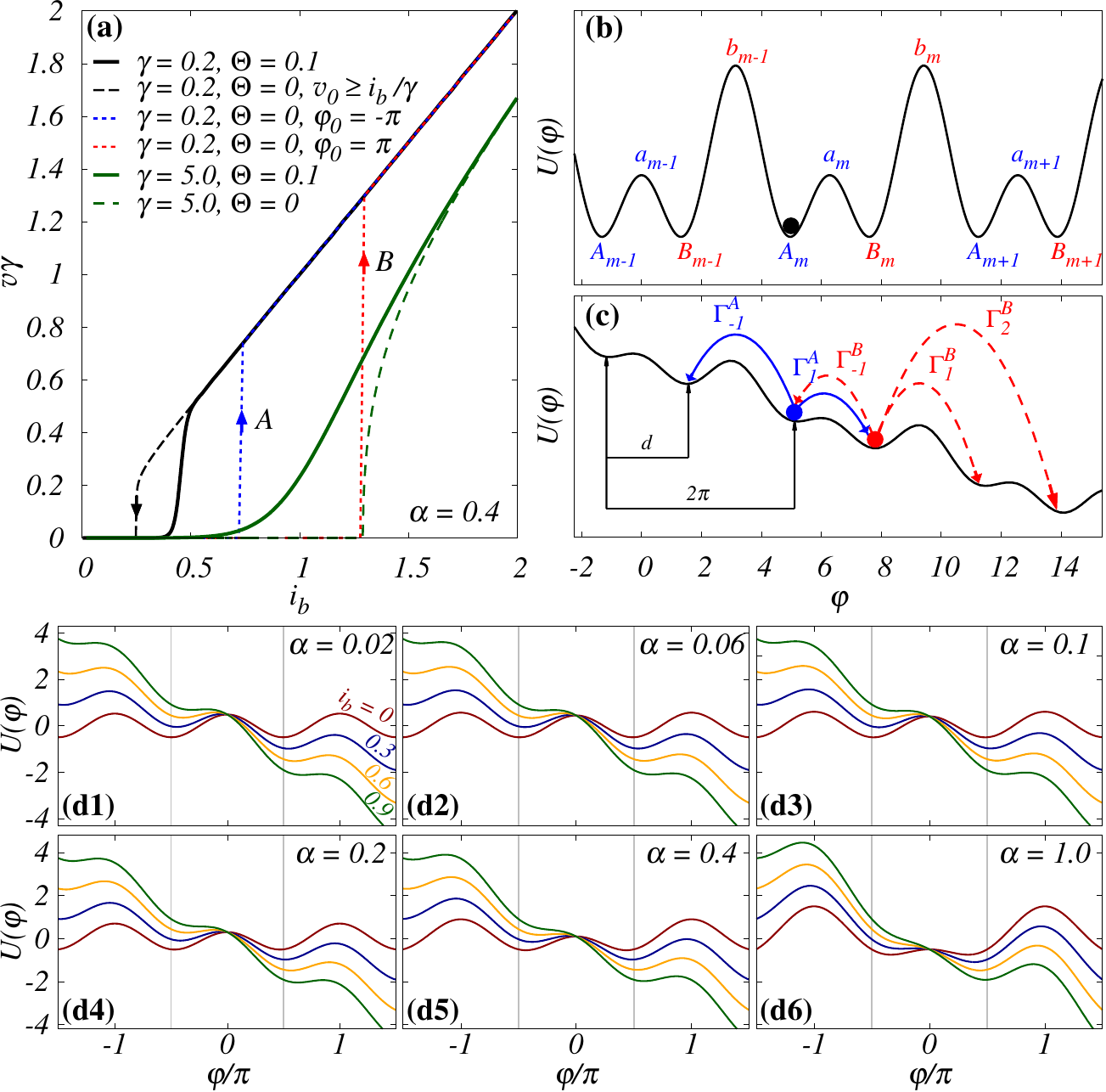}
	\caption{(a) Examples of $v-i_{b}$ characteristics representing the strong
		damping ($\gamma=5$) and the weak damping 
		($\gamma=0.2$) regimes. The solid thick lines show stationary solutions for temperature $\Theta=0.1$. The thin dashed lines show the deterministic (noise-free) solutions. The weak damping case is sensitive to the initial conditions. The black dashed line shows a scenario in which we start with a large force $i_b$, and therefore finite velocity $v$, and then lower $i_b$ slowly enough to reach steady state at each point. The opposite scenario, where we start with $i_b=0$ is represented by a blue dashed line for $\varphi_0$ initially locked in the minimum $A$ and red dashed line for the minimum $B$.  (b) -(c) Illustrations of the potential Eq.~\eqref{eq:Potential1}
with $\Gamma^{A}_{n}$ and $\Gamma^{B}_{n}$ being the rates of the
forward / backward positional jumps over $n$ local maxima. 
(d1)-(d6) Potential profiles for values of $\alpha$ used in the paper and four bias forces $i_b = 0$, $0.3$, $0.6$ and $0.9$. 
The vertical gray lines at $\varphi=\pm\pi/2$ are guides for the eyes that highlight that the minima do not sit at these values and the minima distance $d$ is not necessarily equal to $\pi$.\label{fig:Fig1}}
\end{figure}
\subsection{Model and dynamical regimes}

We consider a stochastic motion of a particle in a potential $U(\varphi)$ with $\alpha\geq0$
described by the dimensionless Langevin equations

\begin{equation}
\begin{split}\frac{\partial v(\tau)}{\partial\tau} & =-\gamma v(\tau)-\frac{\partial U(\varphi)}{\partial\varphi}+\zeta(\tau),\\
v(\tau) & =\partial\varphi/\partial\tau,
\end{split}
\label{eq:Langevin-dimensionless}
\end{equation}
where $v$ is the dimensionless velocity of the particle, $\gamma$ is the friction
coefficient and $\zeta$ represents a Gaussian white noise with the
zero mean $\langle\zeta(\tau)\rangle=0$ and correlation function $\langle\zeta(\tau_{1})\zeta(\tau_{2})\rangle=2\gamma\Theta\delta(\tau_{1}-\tau_{2})$
where $\Theta$ is the dimensionless temperature. 
In what follows, we assume that the system is in the classical regime. This means that $\Theta$ is much higher than some critical $\Theta_c$ of the quantum-classical transition. This critical temperature depends on details of the physical realization of the system and is not part of the effective model in Eq.~\eqref{eq:Langevin-dimensionless} which already assumes $\Theta \gg \Theta_c$. This assumption must be verified for each particular realization of the problem because quantum effects, such as macroscopic quantum tunneling known from Josephson junctions~\cite{caldeira1981influence,grabert1984crossover,massarotti2015macroscopic,Menditto2018Evidence}, can significantly alter the particle dynamics if this condition is not fulfilled.

The associated
Fokker-Planck equation~\cite{Risken} of Eqs.~\eqref{eq:Langevin-dimensionless}
for the probability distribution function $W(\varphi,v,\tau)$ in the
case of potential \eqref{eq:Potential1} reads
\begin{eqnarray}
\label{eq:FullModel}
\dfrac{\partial}{\partial\tau}W(\varphi,v;\tau) & = &
\dfrac{\partial}{\partial v}\left(\gamma v+\alpha\sin\varphi-\sin2\varphi-i_{b}\right)W\label{eq:Fokker-Planck}\nonumber\\
&&-v\dfrac{\partial}{\partial\varphi}W+\gamma\Theta\dfrac{\partial^{2}}{\partial v^{2}}W\\
& \equiv & L_{\mathrm{FP}}W(\varphi,v;\tau).\nonumber 
\end{eqnarray}
The derivation of the average velocity 
\begin{equation}
\langle v\rangle=\int\limits _{0}^{2\pi}\mathrm{d}\varphi\int\limits _{-\infty}^{\infty}\mathrm{d}vvW_{\mathrm{stat}}(\varphi,v)
\end{equation}
follows closely the standard MCF method (see Ref.~\cite[Sec.~9]{Risken} for a general introduction to MCF and Ref.~\cite[Sec.~11.5]{Risken} for the variant here used)
and the derivation of (zero-frequency) velocity noise 
\begin{equation}
S=\int\limits _{-\infty}^{\infty}\mathrm{d}\tau\big(\langle v(\tau)v(0)\rangle-\langle v(\tau)\rangle\langle v(0)\rangle\big)
\end{equation}
in the stationary state $W_{\mathrm{stat}}(\varphi,v)\equiv\lim_{\tau\to\infty}W(\varphi,v;\tau)$
follows that of Ref.~\cite{Zonda15}. 

Figs.~\ref{fig:Fig1}(b)-(c) illustrate the system for zero and
finite bias force, and additional examples relevant for our study are plotted in panels (d1)-(d6). The potential \eqref{eq:Potential1} has for $\alpha<2$
and zero bias two types of minima marked as $A$ and $B$ and two
types of maxima marked as $a$ and $b$. Consequently, there are also two
critical bias forces: $i_{cA}$ where the minimum $A$ and maximum
$a$ disappear and $i_{cB}$ where the minimum $B$ and maximum $b$
disappear. The system can be in three distinct stationary
regimes. Namely, the particle can be running, i.e., at high bias force,
or it can be locked in one of the potential minima types. Consequently, for the noiseless case, the initial position of the particle can, depending on the damping, play an important role even for the steady state, as illustrated in Fig.~\ref{fig:Fig1}(a). 

To explain this, let us consider two limiting cases, one with a strong damping and the other with a weak damping. In both cases, first slowly (compared to any other process) ramp the bias
force from $i_{b}=0$ to $i_{cB}$ with the aim of observing the escape
of the particle from a potential well. Afterwards, we assume a
backward ramping from $i_{b}\gg i_{cB}$ to $i_{b}=0$ with the aim
of observing the recapture (retrapping) of the running particle in
one of the wells. 

For strong damping [$\gamma=5$ in Fig.~\ref{fig:Fig1}(a)]
there is only a single escape bias force and it is identical to
the retrapping force. Due to the strong damping, the initial position
of the trapped particle does not play a role in this regime. The particle
will start running only after the higher maximum $b$ disappears and,
vice versa, will be retrapped at the minimal bias force for which
the maximum $b$ appears. Therefore, both the critical escape force
and the retrapping force are equal to $i_{cB}$ (red dashed line). 

In contrast, there are two possible escape tilts for weak
damping ($\gamma=0.2$). If the particle initially is trapped at the minimum $A$, it will obtain enough inertia to overcome
the still existing maximum $b$ already at the first critical force
$i_{cA}$. However, if it is initially trapped at the minimum $B$ it will
stay there until the second critical force $i_{cB}$ is reached.
The retrapping is also more complicated than in the strong-damping case. If the particle is already running, then it has enough inertia
to overcome the local maxima existing below the critical force $i_{cB}$
or even $i_{cA}$  up to the actual retrapping force
$i_{r}$~\cite{McCumber68}, black dashed lines in Fig.~\ref{fig:Fig1}(a). The retrapping force is determined by
the energy balance between the energy supply of the bias force and
the dissipation~\cite{Risken,Goldobin13}. This means that there is
a region of coexistence of the running and the locked state solutions.

\begin{figure}[t]
 	\includegraphics[width=1\columnwidth]{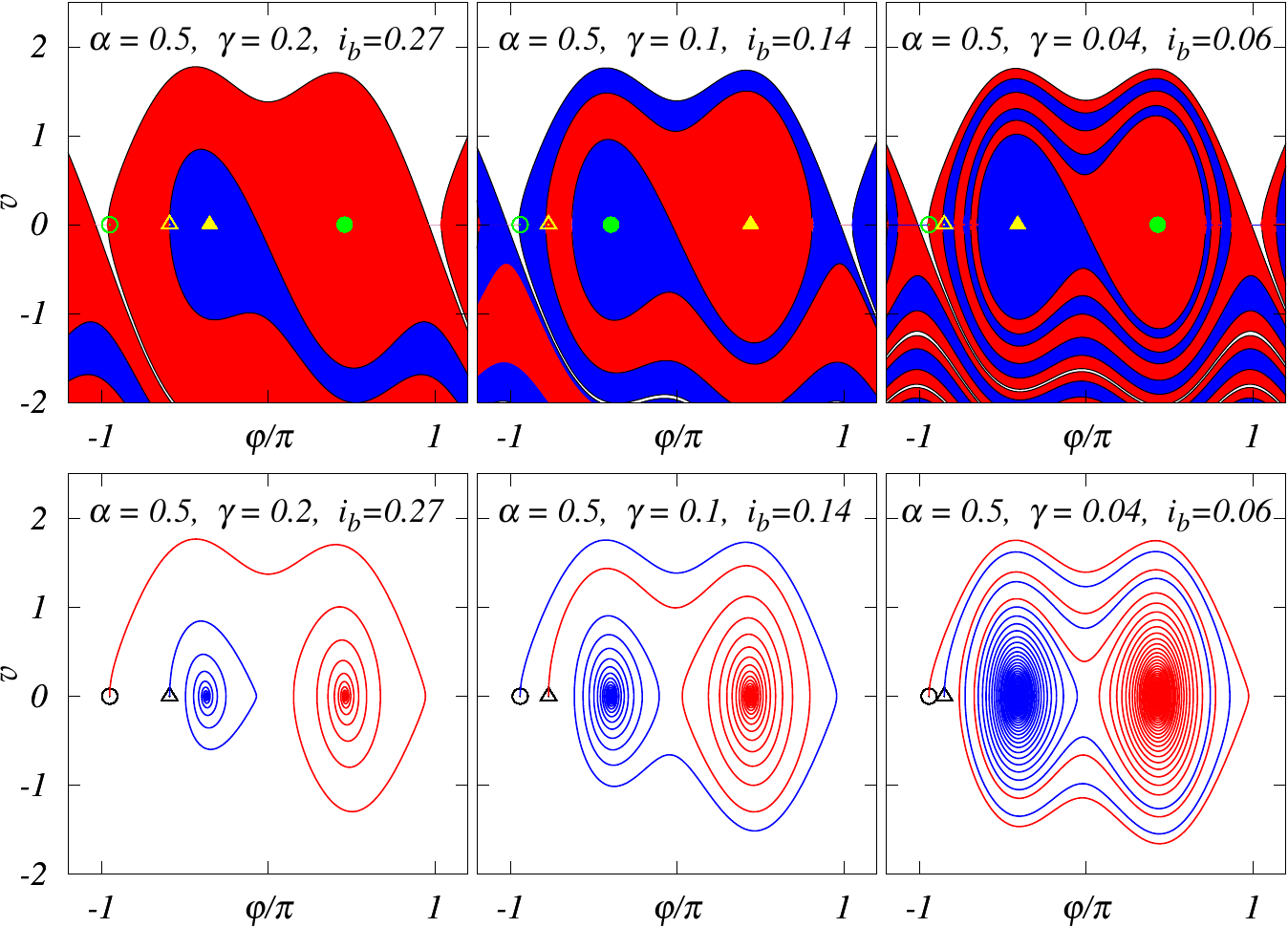}\caption{Top row: Separatrices (black lines) shown for different dampings and bias forces that illustrate the complicated retrapping dynamics of the noiseless case. Any particle within the white region will keep running forever, particles within the blue and red regions will get trapped following paths bounded by the regions of the same color as their initial state (they cannot cross a separatrix). For example, a motionless particle placed initially close to the top of the higher potential maximum, but just right of the running-locked separatrix as marked by the green empty circles ($v=0,$ $\varphi=\varphi^{b}+\delta$), will end in a minimum marked by the green filled circle. Bottom row: Examples of the retrapping trajectories for initial conditions marked by the empty circles and triangles. The blue lines mark the trajectories bounded to the locked solution $A$ and the red lines to the locked solution $B$.\label{fig:Separatrixes}}
\end{figure}

The question of in which minimum the particle will be trapped requires careful analysis because it is a parameter-sensitive process. We illustrate this in Fig.~\ref{fig:Separatrixes} using three sets of parameters for which all three solutions can be realized depending on the initial conditions, with the force $i_b$ being close to the retrapping one. The top row shows examples of separatrix curves and the respective regions of initial conditions that lead to running (white), locked $A$ (blue), and locked $B$  (red) solutions. Separatrices are $2\pi$ periodic and become more complex with decreasing $\gamma$ and $i_b$.  As a consequence, the retrapping trajectory of a particle can be very complicated. This is illustrated in the bottom row of Fig.~\ref{fig:Separatrixes}. The initial conditions are marked by empty symbols (circles or triangles), and the final state is marked by filled ones. The right column illustrates a setting in which even a tiny monotonic change in the initial position of the particle can lead to switching between the two locked final states, that is, a deterministic butterfly effect discussed in more detail, e.g., in Refs.~\cite{Goldobin13,Menditto16_PRB}.   

The noise makes the dynamics even more complicated. Additional processes,
such as switching between running and locked states or occasional
jumps of the particle over the neighboring maxima [see Fig.~\ref{fig:Fig1}(c)],
are possible for nonzero temperature. These processes affect the stationary probability distribution function and, therefore, also the relevant mean values. The solid curves plotted in Fig.~\ref{fig:Fig1}(a)
represent the mean stationary velocity of the particle for temperature
$\Theta=0.1$. As a consequence of the noise, the strong damping
case shows a smooth and shallow crossover from the locked to the running
state, while the weak damping case exhibits a much sharper transition placed between the
retrapping and the lower of the escape bias forces. 

There are three main components that contribute to the overall velocity
noise of this model~\cite{Zonda12,Zonda15}. The thermal noise component dominates close to the equilibrium $i_{b}=0$. The second component is the switching noise
coming from the switching between running and locked states, which is,
for low enough temperatures, an effective dichotomous-like process.
The third component is the shot noise (PJ regime) related to rare jumps
of the particle over single or multiple maxima. The regimes where particular components prevail can be identified from the Fano factor
$F\equiv S/(2\pi\left\langle v\right\rangle)$~\cite{Zonda15}, which
is a normalized noise-to-signal ratio.  

\subsection{Full-counting statistics for the positional jump dynamics}

The positional jumps are, for low enough temperature and weak bias force, well defined distinct (rare) events that significantly influence the overall dynamics. To calculate the rates of these events,
we have adapted the full-counting statistics technique previously used
to study jump probabilities in single-harmonic systems~\cite{Ferrando92,Ferrando93,Golubev10,Zonda15}.
The double-well character of the potential~\eqref{eq:Potential1} requires some generalizations of this method which we present here. 

In the first step we have approximated the solution of the Fokker-Planck equation \eqref{eq:Fokker-Planck} for sufficiently low biases and temperatures by a weighted sum of quasi-equilibrated sharp ($\Theta\ll1$) Gaussian distributions~\cite{Golubev10} around the two types of local minima
\begin{align}
\label{eq:DisAp}
W(\varphi,v;\tau)&\approx\sum_{m}P^{A}_{m}(\tau)w(\varphi-\varphi^{A} _{m},v)\nonumber\\
&+\sum_m P^{B}_{m}(\tau)w(\varphi-\varphi^{B}_{m},v),\\
\text{where: }w(\varphi,v)&=\tfrac{\exp\left(-\varphi{}^{2}/2\Theta\right)\exp(-v^{2}/2\Theta)}{2\pi\Theta}.\nonumber
\end{align}
Here $\varphi^{A}_{m}$ and $\varphi^{B}_{m}$ are
the positions of the $m$-th potential minima and $P^{A}_{m}(\tau)$ and $P^{B}_{m}(\tau)$
are the corresponding time-dependent weights. These are assumed to satisfy
the (Markovian) master equations (MEs) 
\begin{eqnarray}
\frac{dP^A_{m}}{d\tau} & = & \sum_{j}\left(\Gamma^{A}_{2j}P^{A}_{m-j}-\Gamma^{A}_{j}P^{A}_{m}+\Gamma^{B}_{2j-1}P^{B}_{m-j}\right),\label{eq:ME1}\\
\frac{dP^{B}_{m}}{d\tau} & = & \sum_{j}\left(\Gamma^{B}_{2j}P^{B}_{m-j}-\Gamma^{B}_{j}P^{B}_{m}+\Gamma^{A}_{2j+1}P^{A}_{m-j}\right),
\end{eqnarray}
where $\Gamma^{A}_{n}$ is the rate of a positional jump from the potential
well $A$ and $\Gamma^{B}_{n}$ from the well $B$ over $n$ potential local 
maxima to another local minimum. Rates with even $n$ belong to positional jumps between minima of the same kind ($A\rightarrow A$,
$B\rightarrow B$) and odd ones to positional jumps between minima
of different kinds ($A\rightarrow B$, $B\rightarrow A$). The negative
$n$'s correspond to the jump rates in the direction opposite the slope of the bias (up the hill). Following the standard FCS methodology~\cite{Golubev10},
we can evaluate the $k$-dependent cumulant generating function (where $k$ is the counting field) for long times from the ME and equate it with the cumulant generating function of the full model
\begin{equation}
\mathcal{F}(k;\tau\rightarrow\infty)\equiv\ln\int\limits _{-\infty}^{\infty}\mathrm{d}\varphi e^{ik\varphi}\intop\limits _{-\infty}^{\infty}dvW(\varphi,v;\tau\rightarrow\infty),
\end{equation}
calculated by MCF as explained in the Appendix of Ref.~\cite{Zonda15}.
The approximate probability density following from the ME reads
\begin{align} 
\exp\left[\mathcal{F}^{\textrm{PJ}}(k;\tau)\right]=&e^{ik\varphi_{A}}\sum_{m}P^{A}_{m}e^{ik2\pi m}\nonumber\\&+e^{ik\varphi_{B}}\sum_{m}P^{B}_{m}e^{ik2\pi m}\nonumber\\ \equiv&\mathcal{P}_{A}(k,\tau)+\mathcal{P}_{B}(k,\tau).
\end{align}
The probability densities must satisfy the matrix equation
\begin{gather}
\frac{d}{d\tau}\left(\begin{array}{c}
\mathcal{P_{\mathrm{A}}}(k,\tau)\\
\mathcal{P_{\mathrm{B}}}(k,\tau)
\end{array}\right)=\left(\begin{array}{cc}
H_{11}^\textrm{PJ} & H_{12}^\textrm{PJ}\\
H_{21}^\textrm{PJ} & H_{22}^\textrm{PJ}
\end{array}\right)\left(\begin{array}{c}
\mathcal{P_{\mathrm{A}}}(k,\tau)\\
\mathcal{P_{\mathrm{B}}}(k,\tau)
\end{array}\right),\label{eq:MEk}
\end{gather}
where the matrix $\mathbf{H}^\textrm{PJ}(k)$ reads
\begin{gather}
\mathbf{H}^\textrm{PJ}(k)=\left(\begin{array}{cc}
-\sum\limits _{j}\Gamma^{A}_{j}+\Gamma^{A}_{2j}e^{ik2\pi j} & \sum\limits _{j}\Gamma^{B}_{2j-1}e^{ik(2\pi j-d)}\\
\sum\limits _{j}\Gamma^{A}_{2j+1}e^{ik(2\pi j+d)} & -\sum\limits _{j}\Gamma^{B}_{j}+\Gamma^{B}_{2j}e^{ik2\pi j}
\end{array}\right).\
\end{gather}
The $d$ factor in the exponents of the above equations is the
distance between the neighboring minima $d=\varphi^{B}_{m}-\varphi^{A}_{m}$
[see Fig.~\ref{fig:Fig1}(c)] which depends on the potential parameters
$\alpha$ and $i_{b}$. In contrast to even positional jumps, where
the distance traveled is always an integer multiple of $2\pi$,
odd jumps overcome a distance of $2\pi j+d$ where in general $d\neq\pi$,
therefore, we call these jumps \textit{fractional}.
Note, that we use the term \emph{fractional} in its general sense of something that is less than a whole or less than a complete unit and we do not imply that the jumps are necessary rational numbers in the units of $\pi$.

Analogously to Appendix in Ref.~\cite{Zonda15} we use MCF to calculate the two counting-field-dependent eigenvalues $\lambda_{0}(k),\:\lambda_{1}(k)$ with the largest real parts
of the full problem~\eqref{eq:FullModel} with a modified boundary condition $W(\varphi,v;\tau\hspace{-4pt}\shortrightarrow \hspace{-4pt}\infty)=e^{i2\pi k}W(\varphi,v;\tau\hspace{-4pt}\shortrightarrow\hspace{-4pt}\infty)$ and related eigenvectors $u_{0}(k,\varphi,v)$, $u_{1}(k,\varphi,v)$ ~\cite[Sec.~9.3]{Risken}. 
The above two eigenvalues are in the relevant regime well separated from all
the others.

\begin{figure}
	\includegraphics[width=1\columnwidth]{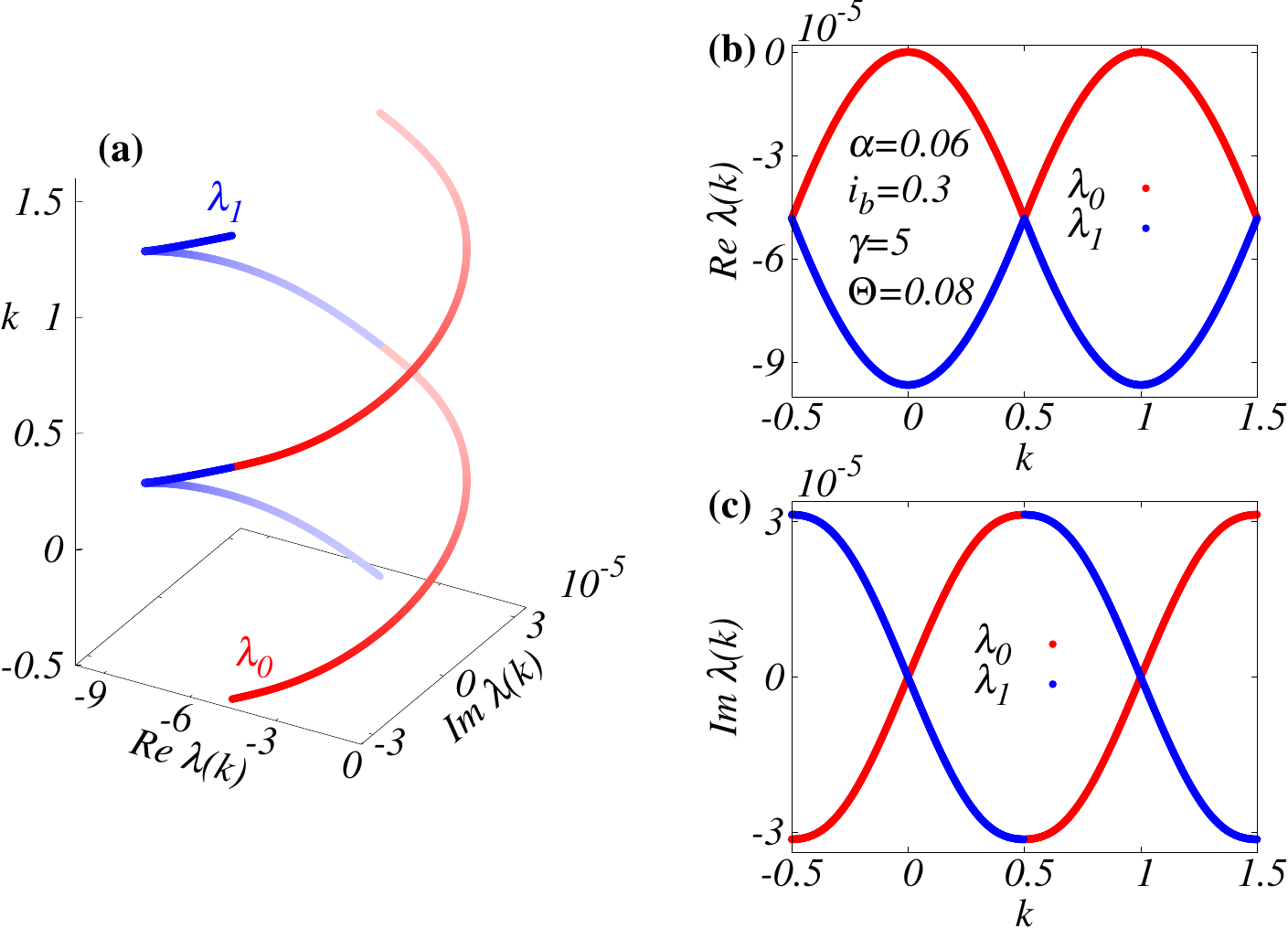}\caption{The two eigenvalues with the biggest real part plotted at the
		Riemann surface (a) and their projections to the real (b) and
		the imaginary (c) planes. \label{fig:Riemann}}
\end{figure}

We construct two component vectors from the eigenvectors by integration over the basins of attraction of the respective nonequivalent local potential minima
\begin{equation}
U_j(k) =  
\begin{pmatrix}
\int_{-\infty}^{\infty}dv\int_{\varphi^b_{m-1}}^{\varphi^a_m}d\varphi u_{j}(k,\varphi,v)\\ 
\int_{-\infty}^{\infty}dv\int_{\varphi^a_{m}}^{\varphi^b_m}d\varphi u_{j}(k,\varphi,v)
\end{pmatrix}.
\end{equation}
They can be used to reconstruct the
matrix $\mathbf{H}^\textrm{PJ}(k)=\mathbf{U}(k)\mathbf{L}^\textrm{PJ}(k)\mathbf{U}^{-1}(k)$ where
\begin{equation}
\mathbf{L}^\textrm{PJ}(k)=\left(\begin{array}{cc}
\lambda_{0}(k) & 0\\
0 & \lambda_{1}(k)
\end{array}\right)\label{eq:eigval}
\end{equation}
and $\mathbf{U}(k)$ is the square matrix of vectors $U_0(k)$ and $U_1(k)$. Note that
due to the second harmonics in the potential and similarly to other non-Hermitian Hamiltonian systems~\cite{Ren2013braid,Li2014quantum}, the eigenvalues~\eqref{eq:eigval} can have complicated topological properties, as shown in Fig.~\ref{fig:Riemann}. The two eigenvalues can, for small enough $\alpha$,
connect at $k_{c}=1/2+j$, where $j$ is an
integer, and smoothly continue each other on the Riemann surface; see Fig.~\ref{fig:Riemann}(a).
Therefore, the real parts of these two eigenvalues, plotted
as functions of $k$, touch on $k_{c}$ [Fig.~\ref{fig:Riemann}(b)], and there is a discontinuity
in the imaginary part of the eigenvalues at the same points [Fig.~\ref{fig:Riemann}(c)].
These discontinuities must be treated with care in the numerical evaluation
of the eigenvalues and eigenvectors. 

Having the reconstructed matrix $\mathbf{H}^\textrm{PJ}$ one can evaluate
the rates of the even positional jumps, i.e., jumps between the same
kind of minima with distance traveled $2\pi j=\pi n$,  using the transformations 
\begin{eqnarray}
\Gamma^{A}_{2j} & = & \intop_{-1/2}^{1/2}H_{11}^\textrm{PJ}e^{-ik2\pi j}dk,\nonumber\\
\Gamma^{B}_{2j} & = & \intop_{-1/2}^{1/2}H_{22}^\textrm{PJ}e^{-ik2\pi j}dk,\label{eq:RatesEven}
\end{eqnarray}
and the odd ones between the different minima types, with distance traveled $\pi (n\mp 1) \pm d$, as 
\begin{eqnarray}
\Gamma^{A}_{2j+1} & = & \intop_{-1/2}^{1/2}H_{21}^\textrm{PJ}e^{-ik(2\pi j+d)}dk,\nonumber\\
\Gamma^{B}_{2j-1} & = & \intop_{-1/2}^{1/2}H_{12}^\textrm{PJ}e^{-ik(2\pi j-d)}dk.\label{eq:RatesOdd}
\end{eqnarray}

This method also provides a simple tool to check its validity.
The approximated mean velocity and velocity noise can be calculated directly from the rates by evaluating the formulas
\begin{align}
\langle v\rangle & =\lim_{\tau\rightarrow\infty}-\frac{i}{\tau}\left.\frac{\partial\mathcal{F}^\textrm{PJ}(k,\tau)}{\partial k}\right|_{k=0},\nonumber\\
S & =\lim_{\tau\rightarrow\infty}-\frac{1}{\tau}\left.\frac{\partial^{2}\mathcal{F}^\textrm{PJ}(k,\tau)}{\partial k^{2}}\right|_{k=0}.\label{eq:FCSvS}
\end{align}
These results can be compared with the mean velocity and overall noise
obtained directly from the MCF method calculations~\cite{Zonda15}.

\section{RESULTS}

\subsection{Strong damping}

We start our analysis with the strong damping case.
The main reason is that in this regime the dynamics is much simpler than for the intermediate
and weak damping. Nevertheless, it is still far from
trivial.


\subsubsection{Fano factor and positional jumps\label{subsec:Fano-factor-and}}

\begin{figure}
	\includegraphics[width=1\columnwidth]{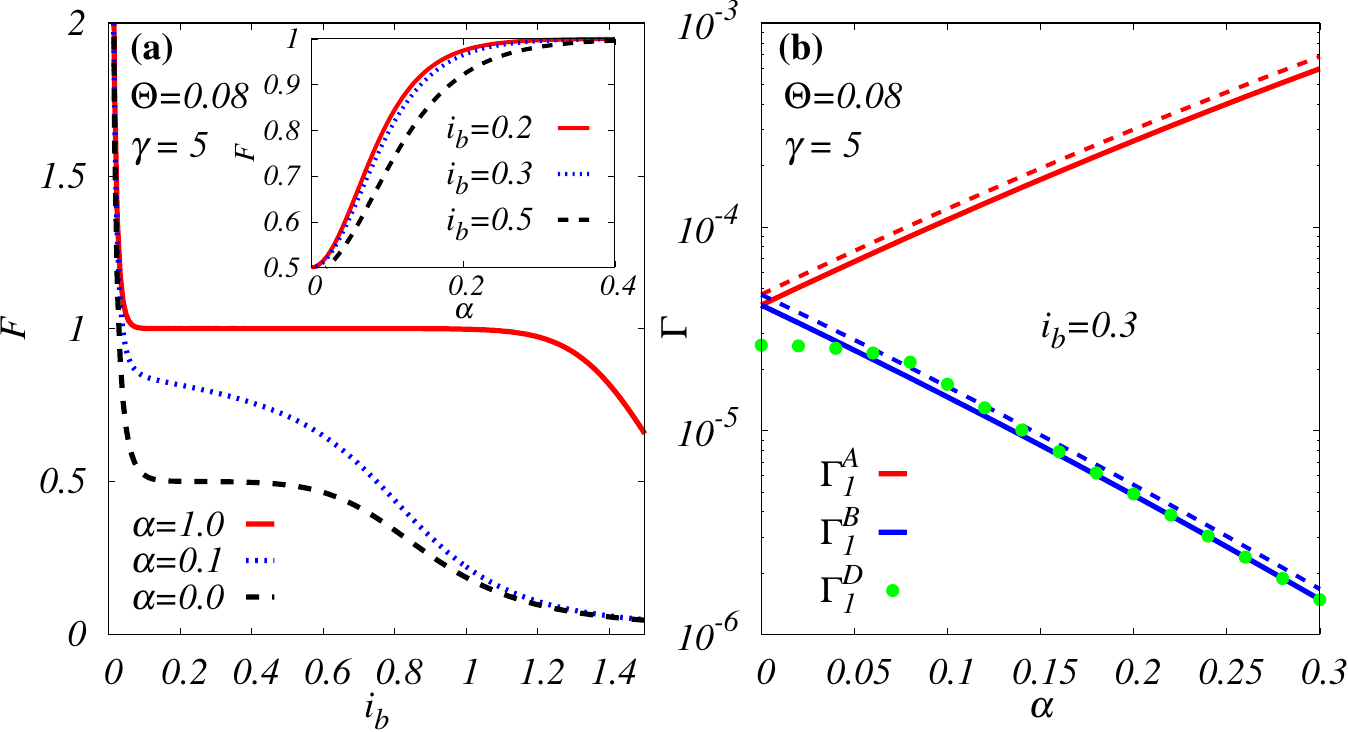}\caption{(a) Examples of typical low-temperature Fano factor $F=S/(2\pi\langle v\rangle)$ for strong
		damping plotted as a function of bias force for different parameters
		$\alpha$. The inset illustrates how the $0.5$ Fano factor plateau transits into Fano plateau with value 1 as $\alpha$ increase. (b) A comparison of the $\alpha$-dependence of the single-positional-jump rates
		evaluated numerically (solid lines) with the analytical Kramers formula result for the overdamped case~\cite{Kramers40,Golubev10} (dashed lines). The green bullets represent rates obtained via the simplified model where only $2\pi n$ PJs were taken into account.\label{fig:StrongDamping}}
\end{figure}

In Fig.~\ref{fig:StrongDamping}(a) we show the Fano factor in the strongly damped regime represented by $\gamma=5$ for three values of
$\alpha$ at temperature $\Theta = 0.08$. The Fano factor exhibits a characteristic divergence close to the equilibrium ($i_{b}=0$) due to the finite thermal noise. For $\alpha=0$, it follows formula $F=\frac{1}{2}\coth(\pi i_{b}/2\Theta)$~\cite[Eq.~(28)]{Golubev10} for small $i_{b}$ and $\alpha=0$ with
a plateau at the Poissonian value of $F=0.5$ (note that we still normalize
$F$ to $2\pi$). This is in agreement with the dominant contribution to the noise from the single PJ [illustrated
in Fig.~\ref{fig:Fig1}(c)] over a distance $d=\pi$ between
the equivalent neighboring minima. The situation for the finite $\alpha$
is more complicated. The range in which single PJs are the dominant
source of the overall noise is not marked by a clear plateau.
Rather for $0<\alpha\ll1$ [$\alpha=0.1$
in Fig.~\ref{fig:StrongDamping}(a)] we observe a slight slope
in the Fano factor. This reflects the fact that for finite $\alpha$
the prevailing velocity noise contribution consists of a nontrivial
combination of the single positional jumps forward over the distance of $\varphi^{B}_{m}-\varphi^A_m=d$
and backward by $\varphi^{A}_{m+1}-\varphi^B_{m}=2\pi-d$ [see Fig.~\ref{fig:Fig1}(c)]
where $d$ depends on both $\alpha$ \emph{and} $i_{b}$. As already stated, because $d\neq\pi$ we refer to these events as the \textit{fractional positional jumps}. 

With increasing $\alpha$ the Fano factor increases smoothly in this region from $0.5$ signaling only PJ over the distance $\pi$ to $1$ suggesting a single PJs over $2\pi$ as shown in the inset
of Fig.~\ref{fig:StrongDamping}(a). Interestingly, for $\Theta=0.08$ the Fano factor approaches 1 even for values of $i_{b}$ that are still well below $i_{cA}$, therefore, in a regime where both minima still exist. For example, the critical current $i_{cA}$ is approximately $0.396$ at $\alpha = 1$, yet its Fano factor follows the analytical result $F=\coth(\pi i_{b}/\Theta)$, valid for systems with only the first harmonic, even below this value. Nevertheless, this can be understood as a consequence of the large $\gamma$ and can be explained using the FCS method together with a simplified model of the elementary PJ processes.

Because only jumps over a single (uneven) maxima are realized in the locked state of the strong-damping case the
overall dynamics of this regime can be described using just four rates:
$\Gamma^{A}_{1}$ and $\Gamma^{B}_{1}$ for the forward single jumps
and $\Gamma^{A}_{-1}$ and $\Gamma^{B}_{-1}$ for the backward single
jumps. Moreover, the backward rates can be neglected if the bias force
is strong enough. The typical dependencies of $\Gamma^{A}_{1}$ and
$\Gamma^{B}_{1}$ on $\alpha$ in this regime represented
by the bias force $i_{b}=0.3$ and the damping $\gamma=5$ are plotted
in Fig.~\ref{fig:StrongDamping}(b) (solid lines). Both $\Gamma^{A}_{1}$
and $\Gamma^{B}_{1}$ closely follow the Kramers formula for escape
across the adjacent barrier for overdamped case $\Gamma^{X}_1=\frac{1}{2\pi}\sqrt{|U''(\varphi^{x})|U''(\varphi^{X})}e^{-\Delta U_{X}/\Theta}$~\cite{Kramers40,Golubev10} (where $\Delta U_{X}=U(\varphi^{x})-U(\varphi^{X})$, and $x=a,b$; $X=A,B$)
plotted with dashed lines of the respective colors. Note that because of the increasing difference
$\Delta U_{B}-\Delta U_{A}$
the ratio
\begin{equation}
\Gamma^{A}_{1}/\Gamma^{B}_{1}\sim\exp\left[\left(\Delta U_{B}-\Delta U_{A}\right)/\Theta\right]\label{Eq:Ratio}
\end{equation}
increases exponentially with $\alpha$. Consequently, for high enough $\alpha$
and low enough $\Theta$ the average waiting time for the escape from minimum $A$ $\tau_{A\rightarrow B}=1/\Gamma^{A}_{1}$ is
negligible compared to
the waiting time for the escape from minimum $B$ ($\tau_{B\rightarrow A}=1/\Gamma^{B}_{1}$).
Therefore, in the long term, every single positional jump $B_{m-1}\rightarrow A_{m}$ is
immediately followed by a single positional jump $A_{m}\rightarrow B_{m}$.
The combination of these two fractional PJs is effectively a complete
$2\pi$ jump. This is shown in Fig.~\ref{fig:StrongDamping}(b)
by the green bullets that were obtained via a simplified model where only the
$2\pi n$ jumps were considered~\cite{Zonda15}
\begin{equation}
\Gamma^{D}_{n}=\intop\limits _{-1/2}^{1/2}\lambda_{0}(k)e^{-2\pi ikn}dk.\label{eq:GammaDn}
\end{equation}
Their match with the $\Gamma^{B}_{1}$ rate for high enough $\alpha$
is consistent with the Fano-factor value of $1$ in the inset of Fig.~\ref{fig:StrongDamping}(a).
This has interesting physical consequences. If the temperature is
low enough, then, because $\Theta$ is in the denominator of the exponent
of Eq.~\eqref{Eq:Ratio}, any strongly damped $\varphi$ junction or
another equivalent system describable by a double harmonic potential with a finite bias $i_b$
will in the steady state resemble a simple single harmonic system. As such, it can be described by the analytical formulas derived for the single-harmonic potential.   

\subsection{Intermediate and weak damping}

The stochastic dynamics of the particle in the tilted double-well
periodic potential in the regime of intermediate and weak damping is
significantly richer than in the strong damping case. For example, for weak bias force, there are
positional jumps over multiple maxima, and for stronger bias, a complicated switching between running and locked solutions is the prevailing source of the velocity noise. Even the retrapping of a particle from the running to locked regime has complex dynamics, as discussed below.

\begin{figure}
	\includegraphics[width=1\columnwidth]{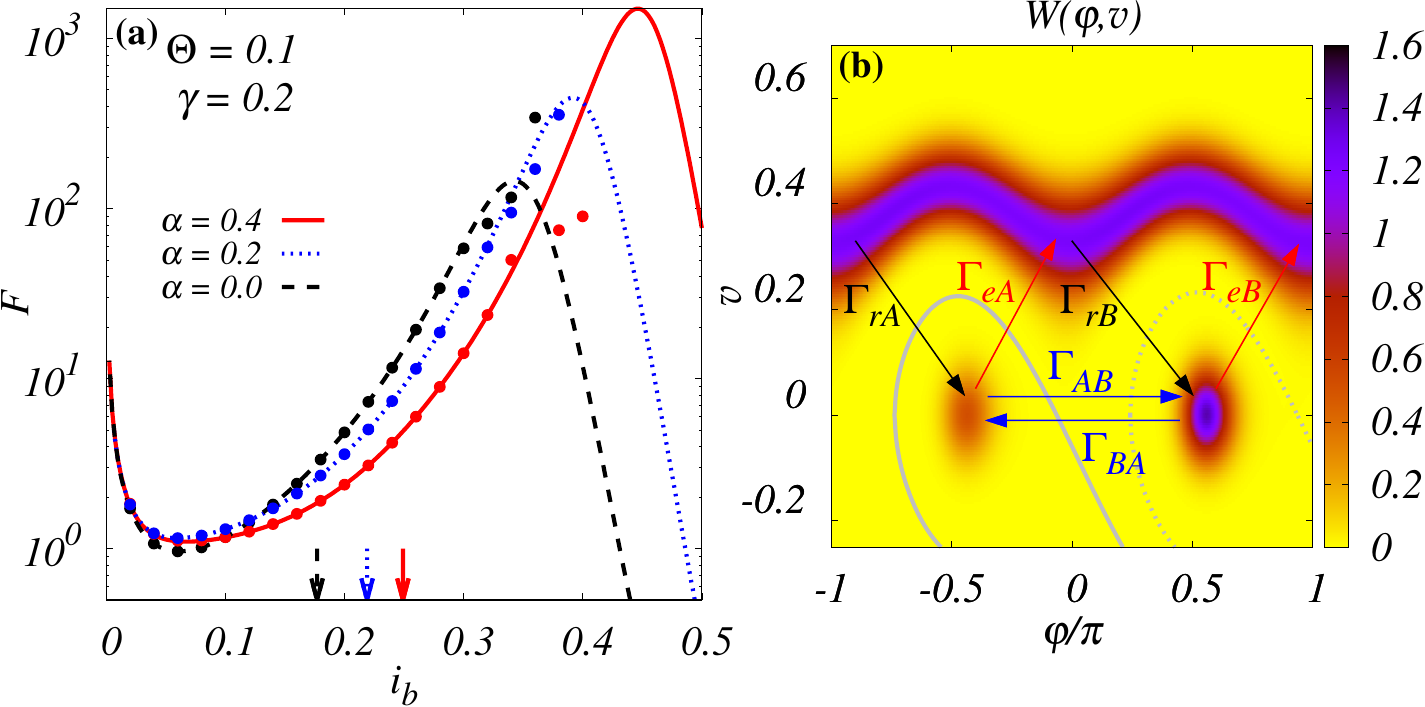}
	\caption{(a) Three examples of a typical low-temperature Fano factor $F=S/(2\pi\langle v\rangle)$ dependence on the bias force for weak damping. The
lines were calculated by the MCF method, and the bullets show the ME results
from Eq.~\eqref{eq:FCSvS}. Vertical arrows
mark the related noiseless retrapping forces. (b) An illustration of the stationary
distribution function $W(v,\varphi)$ in the switching regime calculated for $\gamma=0.2$, $\alpha=0.02$, $i_b=0.38$, and $\Theta=0.08$. The arrows illustrate the processes and their rates as used in Eq.~\eqref{eq:ME2b}.
		Namely, the red arrows represent escape from the locked states to the running one,
the black arrows represent retrapping of the particle, and the blue arrows are the positional jumps between the locked states $A$ and $B$. The gray dashed curves show the separatrices between the running solution and the locked solution in the well $A$ (solid line) and/or well $B$ (dashed line), respectively. \label{fig:Q5FannoMap}}
\end{figure}

The dependencies $F-i_{b}$ for the underdamped case [plotted in Fig.~\ref{fig:Q5FannoMap}(a)] differ qualitatively from the strongly damped case [Fig.~\ref{fig:StrongDamping}(a)]. The dominant feature
of the Fano factor is a huge peak [note the logarithmic scale in Fig.~\ref{fig:Q5FannoMap}(a)] for finite $i_b$.
As was shown in our previous study of the RCSJ model with single
harmonics CPR~\cite{Zonda15}, this peak is a consequence of the switching
process between coexisting, but well separated, running and locked
states in this range of $i_b$. This interpretation is also valid for the double-well
potential, where, however, there exist two locked states as illustrated in Fig.~\ref{fig:Fig1}.    

We support this claim in Fig.~\ref{fig:Q5FannoMap}(b) where
an example of the stationary distribution function $W(v,\varphi)$ is plotted
for the parameters (see figure description) close to the maximum of the peak. 
Here, the two distinct
peaks at $v\approx0$ centered around potential minima signal the two locked states. The continuous ridge that spreads above them is the running state. 
In this regime, the prevailing contribution to the overall velocity noise comes from the switching between these
three well-separated metastable states. 

When we lower $i_b$ to the regime in between the thermal divergence at $i_b\to 0$ and the switching maximum, multiple fractional positional jumps become the prevailing source of the velocity noise. We now analyze the regimes of positional jumps and switching separately and then show how they transition smoothly into each other.  

\subsubsection{Positional jumps\label{subsec:Phase-jumps}}

\begin{figure*}
	\includegraphics[width=1\textwidth]{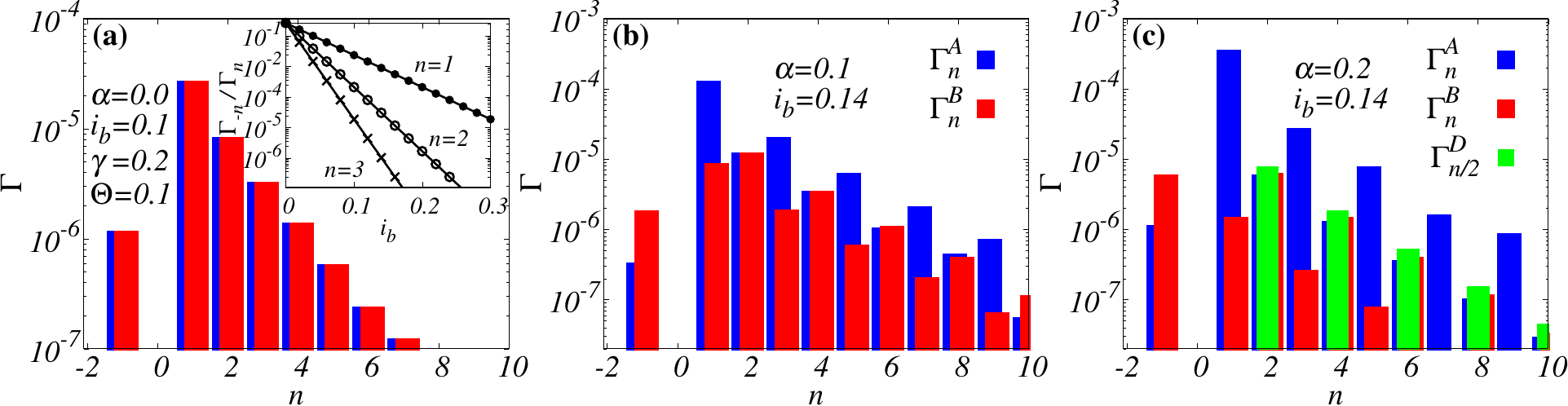}\caption{The rates of multiple fractional positional jumps (of order $n$) for different values of the bias force $i_{b}$ and ratio $\alpha$. The inset in (a) is a verification
of the detailed balance condition. The blue columns describe jumps starting at minimum $A$, red at minimum $B$ (a small horizontal shift is introduced for visibility), and the green columns in panel (c) show the rates
obtained via the simplified model where only $2\pi n$ PJs were considered [analogously to Fig.~\ref{fig:StrongDamping}(b)].\label{fig MPS}}
\end{figure*}

In Fig.~\ref{fig MPS} we show the rates of multiple positional
jumps for weak damping $\gamma=0.2$, low temperature $\Theta=0.1$,
and different values of $\alpha$ and $i_{b}$, where $n$ is the
number of maxima bridged by a single process. The blue columns represent
the rates of jumps from the minima $A$ and the red ones from the minima $B$. The $n$ dependence of the PJ rates for $\alpha=0$
shown in panel (a) is the same as that for the simple single-harmonic potential~\cite{Zonda15} up to a redefinition of the jump length.
We use it as one of the tests of our extended FCS method. 
In the inset of Fig.~\ref{fig MPS}(a) we show the ratios $\Gamma_{-n}/\Gamma_{n}$ for $n=1,\,2,\,3$
obtained numerically with FCS (bullets and crosses) and analytically
from the detailed balance condition $\Gamma_{-n}/\Gamma_{n}=\exp(-\pi ni_{b}/\Theta)$
with the potential drop of $\pi i_{b}n$ along the positional jumps (lines).
There is a perfect match over seven orders of magnitude for each $n=1,2,3$ proving the reliability of the FCS plus MCF method. 

The profile of the rates for potentials with
finite $\alpha$ is more complicated. The $n$ dependencies of the PJ rates differ qualitatively between jumps starting in different minima, as well as between odd and even $n$'s. We show this in Figs.~\ref{fig MPS}(b)-(c). The rates $\Gamma^A_n$ and $\Gamma^B_n$ differ by several orders of magnitude for odd $n$'s [see $\alpha=0.2$ case in Fig.~\ref{fig MPS}(c)], but are comparable for even $n$'s. This can be rationalized by analyzing the ``trajectories'' of the particular PJs following the illustrations in Fig.~\ref{fig:Fig1}(b)(c). 

The jumps $A_{m}\rightarrow A_{m+n/2}$
and $B_{m}\rightarrow B_{m+n/2}$, where $n$ is an even integer, are indeed comparable. Here, the particle
had to overcome the same number of maxima $a$ as well as $b$, namely $n/2$. In the idealized case, they traveled the same distance $\pi n$. Therefore, also the related rates are equivalent.
However, this is not true for odd jumps representing fractional
PJs between minima of different kinds. In the case of an odd $n$ the particle overcomes $(n+1)/2$ of the lower maxima but only $(n-1)/2$ of the higher ones
during a $\Gamma^{A}_{n}$ process. The opposite is true for the $\Gamma^{B}_{n}$ process. Furthermore, travel distances differ by
$2\pi-2d$. Consequently, the ``odd'' rates differ significantly
($\Gamma^{A}_{n}\gg\Gamma^{B}_{n}$ for positive $n$). 

Taking into account this dynamics, the result that it is more probable
for a particle to travel over $n+1$ maxima than just over $n$
[for example, $\Gamma^{A}_{3}>\Gamma^{A}_{2}$ and $\Gamma^{B}_{2}>\Gamma^{B}_{1}$
in Fig.~\ref{fig MPS}(b),(c)] seems rather paradoxical. However,
this is a problem of ``retrapping'' of a particle. Basically, if the
particle has already overcome the higher maximum $b$, it will also obtain enough inertia in this regime to overcome the next lower maximum $a$. This is a parameter-dependent process, and the retrapping
scenario can be rather complicated~\cite{Goldobin13,Menditto16_PRB}
as is also discussed below in Sec.~\ref{subsec:Retrapping}. 

If the single-jump rate $\Gamma^{A}_{1}$ (slip over single smaller maxima) is much
larger than the rate of any other process, as is typical for high
$\alpha$ and high bias force [Fig.~\ref{fig MPS}(c)], it is again possible
to capture the dynamics of the system using the simplified model described by Eq.~\eqref{eq:GammaDn}.
This is shown in Fig.~\ref{fig MPS}(c), where the green columns represent
the rates of the jumps $2n\pi$, where the double-well character of
the potential is ignored. This model agrees well with the
even rates, which do not reflect the difference between the two kinds of minima. The underlying reason is that the waiting time for an escape from the minima of type $A$ is negligible compared to other time scales.  

Before moving to the switching regime, it is worth stressing that the FCS method works well up to surprisingly
high Fano factors ($F\sim10^{2}$). This is shown in Fig.~\ref{fig:Q5FannoMap}(a)
where the bullets represent the Fano factor obtained directly from
the rates by Eq.~\eqref{eq:FCSvS}.
They are aligned with the curves obtained by the full MCF method up to
the values of $i_{b}$ that are higher than the retrapping forces
of the noiseless scenario [marked with the arrows at the bottom of
Fig.~\ref{fig:Q5FannoMap}(a)]. The mathematical explanation of this
agreement is that for $i_b$ between the positional-jump regime and the switching
regime, the first two eigenvalues with the largest real
parts are still sufficiently separated from the next ones. 
In addition, the comparison in Fig.~\ref{fig:Q5FannoMap}(a) also shows that the multiple positional jumps smoothly change into the running state as we enter the switching regime. Nevertheless, a different approach is needed
to investigate the dynamics in this regime.

\subsubsection{Switching processes\label{subsec:Switching-processes} }
The positional jumps within the double well play an important role even in the regime of larger
bias forces, where the switching between running and locked solutions
takes place. To show this and to evaluate the escape and retrapping rates between locked and running metastable states, we again introduce a simplified model. We divide the full probability distribution
function into three well-separated regions and calculate the occupation probabilities for each of them. The sharp borders between the regions are defined by separatrices of the noiseless ($\Theta=0$) dissipative
steady-state solution. In plain words, we determine the momentary regime of the particle by identifying the deterministic steady state in which the particle would end with its current position and velocity for the noiseless case.  

A trivial example of such a division is plotted in Fig.~\ref{fig:Q5FannoMap}(b).
There, the solid gray curve separates
the $A$-valley locked state with the associated time-dependent occupation
probability $P_{A}(\tau)$, the dashed curve separates the $B$-valley
locked state with the associated occupation probability $P_{B}(\tau)$
and, consequently, the rest of the area belongs to the running state
with occupation probability $P_{R}(\tau)$. The associated occupation
probabilities are assumed to satisfy the master equation
\begin{align}
\frac{d}{d\tau}\left(\begin{array}{c}
P_{R}\\
P_{A}\\
P_{B}
\end{array}\right)=\mathbf{M}^\textrm{S}\left(\begin{array}{c}
P_{R}\\
P_{A}\\
P_{B}
\end{array}\right),\label{eq:ME2}
\end{align}
with the rate matrix
\begin{align}
&\mathbf{M}^\textrm{S}=\nonumber\\
&\left(\begin{array}{ccc}
-\left(\Gamma_{rA}+\Gamma_{rB}\right) & \Gamma_{eA} & \Gamma_{eB}\\
\Gamma_{rA} & -\left(\Gamma_{eA}+\Gamma_{AB}\right) & \Gamma_{BA}\\
\Gamma_{rB} & \Gamma_{AB} & -\left(\Gamma_{eB}+\Gamma_{BA}\right)
\end{array}\right),\label{eq:ME2b}
\end{align}
where $\Gamma_{eA}$ ($\Gamma_{eB}$) is the escape rate
from potential well $A$ ($B$) to the metastable running state; $\Gamma_{rA}$ and $\Gamma_{rB}$ are the retrapping rates from metastable
running state to the potential well $A$ or $B$, respectively, and $\Gamma_{AB}$ ($\Gamma_{BA}$) is the total rate of the
positional jumps (of any length) from potential well $A$ to $B$ ($B$ to $A$) as illustrated in Fig.~\ref{fig:Q5FannoMap}(b).

These rates can be obtained by reconstructing the switching matrix
$\mathbf{M}^\textrm{S}=\mathbf{P}\mathbf{L}^\textrm{S}\mathbf{P}^{-1}$, with the diagonal matrix 

\begin{equation}
\mathbf{L}^\textrm{S}=\left(\begin{array}{ccc}
\lambda_{0} & 0 & 0\\
0 & \lambda_{1} & 0\\
0 & 0 & \lambda_{2}
\end{array}\right)\label{eq:L2}
\end{equation}
containing the three eigenvalues with the largest real parts of the full Fokker-Planck operator calculated via MCF~\cite{Risken}. They represent the stationary state
$\lambda_{0}=0$ and the first two excited states $\lambda_{1}$ and $\lambda_{2}$.
The matrix $\mathbf{P}$ is a square matrix of the three related eigenvectors.
The three components of each eigenvector are obtained by integrating
the full MCF eigenfunctions over the areas bounded by the separatrix
curves of particular solutions. The separatrices are $2\pi$
periodic and, as illustrated in Fig.~\ref{fig:Separatrixes}, can be quite complex when $\gamma\to 0$.  As a consequence, dynamical processes, such as the retrapping of the particle, can be very complicated~\cite{Goldobin13}.

\begin{figure}
	\includegraphics[width=1\columnwidth]{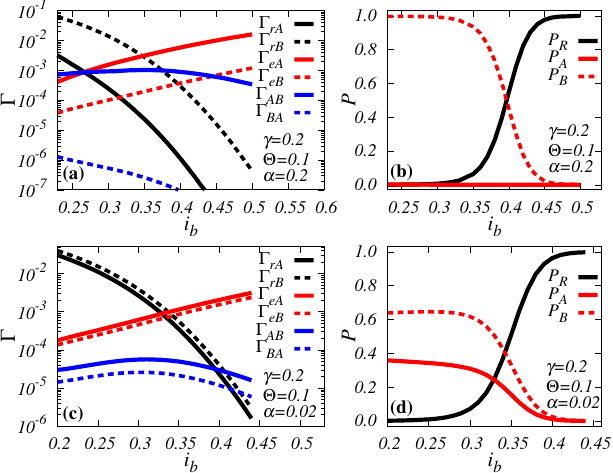}	
	\caption{(a),(c) Dependencies of the escape, retrapping, and positional jump rates [illustrated in Fig.~\ref{fig:Q5FannoMap}(b)] on the bias force within the switching regime calculated using the simplified model~\eqref{eq:ME2}. (b),(d) Related steady-state occupation probabilities. Two qualitatively different cases are presented: panels (a) and (b) show $\alpha=0.2$ where in the steady state the system behaves effectively as a single well, and panels (c) and (d) show $\alpha=0.02$ where the double-well character is evident.\label{fig:Rates} }
\end{figure}

A typical example of the rate dependencies on the bias
force and related stationary occupations of particular states for $\alpha=0.2$ and $\alpha=0.02$ are plotted in Fig.~\ref{fig:Rates}. 
They represent two different regimes.

In panel (a), where $\alpha=0.2$ and $\Theta=0.1$, the positional jump from minimum $A$ to minimum $B$ has a rate $\Gamma_{AB}$ (solid blue
line) that is of the same order
as the escape rate $\Gamma_{eA}$ (solid red line) and much higher
than the escape rate $\Gamma_{eB}$ (dashed red line) at small bias. The rate of the opposite positional jump from $B$ to $A$ (blue dashed line) is negligible. 
In addition, the retrapping rate $\Gamma_{rB}$ (dashed black line) is almost two orders of magnitude larger
than retrapping rate $\Gamma_{rA}$ (solid black line) in the entire plotted range.
This means that the mean lifetime of a particle trapped in the
locked state $A$ is negligible compared to the mean lifetime of the two other
states. Consequently, in the steady state, the system exhibits bistable behavior reminiscent of a system governed by a simple single-harmonic potential. This is also evident in Fig.~\ref{fig:Rates}(b), where the stationary occupation probabilities are plotted. 
For low bias forces, the particle is predominantly trapped in the state $B$, while a sharp transition to the running state is observed near the Fano factor maximum (Fig.~\ref{fig:Q5FannoMap}, blue curve). Throughout the range depicted in the figure, the locked state $A$ exhibits minimal influence on the system dynamics.

\begin{figure}
	\includegraphics[width=1\columnwidth]{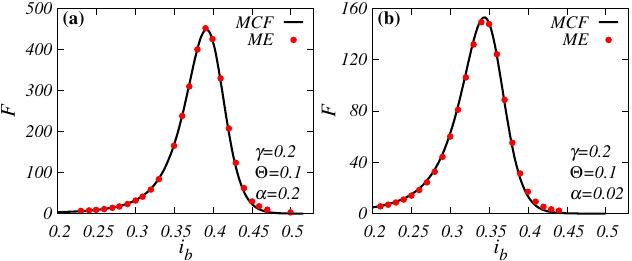}
	\caption{Comparison of the factor factor $F=S/(2\pi\langle v\rangle)$ dependencies on the bias force where the black lines were calculated by the MCF method, and the red circles show the ME results
from Eq.~\eqref{eq:Fme}.
\label{fig:fano_from_ME}}
\end{figure}
 
The situation is qualitatively different for $\alpha=0.02$. In this case, $\Gamma_{AB}$ and $\Gamma_{BA}$ are comparable and considerably smaller than those of the escape and retrapping processes within the range where switching occurs. However, this does not mean that positional jumps between the two locked states are irrelevant. As we discuss later in detail, it only means that their influence is apparent only on much longer timescales. At shorter ones, the dynamics is governed by the escape and retrapping processes. Due to the small $\alpha$, the rates of escape and retrapping processes are comparable ($\Gamma_{rA}\sim\Gamma_{rB}$ and $\Gamma_{eA}\sim\Gamma_{eB}$). Together, this leads to steady-state occupation probabilities where both locked states are relevant, as illustrated in Fig.~\ref{fig:Rates}(d). Interestingly, even with the low value of $\alpha=0.02$, there remains a significant difference between the occupation probabilities of states $A$ and $B$ in the steady state.

The results of the simplified model~\eqref{eq:ME2} can also be used to test whether the peak in the Fano factor actually arises as a result of switching between the running and locked states. In the vicinity of the peak where $F\gg 1$ we can neglect the noise contributions inherent to particular metastable states~\cite{Flindt05,Zonda15} and calculate $F$ from the average velocity (obtained here from MCF) and rates and probabilities from the ME~\eqref{eq:ME2} as~\cite{Zonda15}
\begin{equation}
F_{\mathrm{ME}}=\frac{\langle v \rangle\left(v_r - \langle v \rangle\right)}{\pi v_r \Gamma_e},
\label{eq:Fme}
\end{equation}
where we approximate the velocity in the running state by $v_r=i_b/\gamma$. The total escape rate $\Gamma_e$ is calculated as $\Gamma_e=\bar{P}_{A}\Gamma_{eA} + \bar{P}_{B}\Gamma_{eB}$ where $\bar{P}_{A,B}=P_{A,B}/(P_A+P_B)$ reflect the probabilities with which the already locked particle is placed in minima $A$ or $B$. 
We show in Fig.~\ref{fig:fano_from_ME} a comparison of the Fano factors obtained from Eq.~\eqref{eq:Fme} (red circles) with the full MCF solution (black lines) for the same parameters as used in Fig.~\ref{fig:Rates}. 
The Fano factors are in a very good agreement confirming both the origin of the peak from the switching processes and the validity of the simplified model \eqref{eq:ME2}.

In addition to steady state, the simplified model~\eqref{eq:ME2} allows us to easily investigate the time evolution and to understand its particular time regimes. This can be useful not only by itself but also as a supporting tool for full-scale Langevin simulations. Let us illustrate this by investigating a retrapping process after a parameter quench.    

\begin{figure}[b]
	\includegraphics[width=1.0\columnwidth]{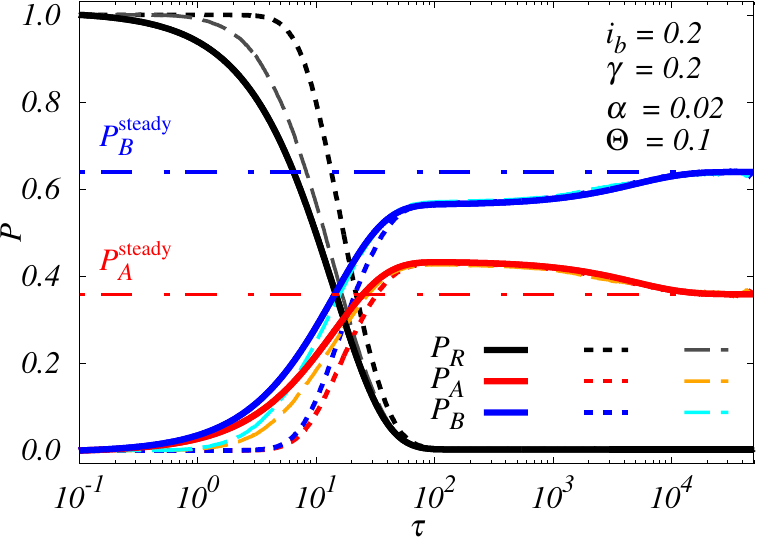}	
	\caption{Example of the time dependence of the occupation probabilities of the three metastable states for $i_b=0.2$ with initial condition $P_R=1$, $P_A=0$ and $P_B=0$. This models a sudden quench of very large bias force to bias force just above the retrapping force of the noiseless case. Solid lines are results of the simplified model~\eqref{eq:ME2}, dashed lines show full Langevin simulations for two different initial conditions.     
		\label{fig:ProbTquench}}
\end{figure}

\subsubsection{Retrapping\label{subsec:Retrapping} }

It is often difficult to reach the steady state in experimental realizations or realistic Langevin simulations. This is especially problematic for systems with low noise and weak damping. However, once we have the rates of the most relevant processes, we can investigate even long-time processes via the master equation~\eqref{eq:ME2}.   

As an illustration, we analyze the scenario of a sudden quench of the parameters.
We employ the master equation~\eqref{eq:ME2} as well as large-scale Langevin simulations. For the latter, we used the modified Euler-Heun method \emph{LambaEulerHeun} from the Julia package \emph{DifferentialEquations.jl} with adaptive integration step. In the simulations, we used $\approx 6.5\times 10^4 $ particles and, as before, their immediate regime was identified via the separatrices for the model parameters after the quench.     

We focus on the case with the steady state illustrated in Fig.~\ref{fig:Rates}(d), i.e., $\alpha=0.02$, $\gamma=0.2$  and $\Theta=0.1$. We prepare the system in a state where it is fully running, for example, with a high bias force. For the simplified model, this means setting $P_R(\tau=0)=1$ and $P_A(\tau=0)=P_B(\tau=0)=0$. For the Langevin simulation, the system was first thermalized at $i_{b}=0.5$, which also gives $P_R(\tau=0)\approx 1$, as evident from Fig.~\ref{fig:Fig1}. In the next step, we abruptly change the bias force to $i_b=0.2$, which is just below the actual retrapping force $i_{r}$. For $i_b=0.2$ the system is already almost exclusively in the locked steady state, yet the running state is still possible and can be identified by its separatrix. 

The calculated time dependencies of the occupation probabilities are shown in Fig.~\ref{fig:ProbTquench} by the solid (master equation) and dashed (Langevin dynamics) black ($P_R$), blue ($P_A$) and red ($P_B$) lines. The logarithmic time scale reveals several metastable regimes. At short times, there are the largest differences between the simplified model and the full simulation, as expected. In particular, the change in initial occupations is faster for the simplified model than in the simulation. This is due to the details of the initial state, that is, the initial distribution of the particles. Full dynamics starts with a relatively high mean velocity of the particles, and therefore it takes a while to slow them down. 

To illustrate that this is indeed the case, we also show a simulation with different initial conditions. The dashed lines in dark gray ($P_R$), cyan ($P_A$) and orange ($P_B$) started in a steady state with $i_b=0.3$ and $\Theta=0.01$ and, therefore, significantly lower mean velocity.
The short-time dynamics of this case approaches that of the simplified model.
Nevertheless, the simplified model almost perfectly predicts, and, more importantly, explains, the full dynamics of these simulations at longer times. 

First, particles are quickly caught in the minima around $\tau\sim10^{1}$, because the retrapping rates are $\Gamma_{rA}\sim\Gamma_{rB}\approx10^{-1}$. The system is close to being completely trapped before $\tau\sim10^{2}$. However, although relatively stable, the occupation probabilities of the locked states are far from their steady-state values (horizontal dash-dotted lines) obtained by the MCF method discussed above. This metastable state exists due to the large separation of the retrapping rates from the escape and positional jumps rates; see Fig.~\ref{fig:Rates}(c). Probabilities start to approach the true steady state only for $\tau\gtrsim10^{4}$
where first the escape processes with rates $\Gamma_{eA}\sim\Gamma_{eB}\sim10^{-4}$
and then positional jumps with $\Gamma_{AB}\sim0.5\times10^{-4}$ start to be relevant. 

This difference of more than two orders of magnitude between the time
of retrapping and reaching the steady state, as well as the existence of the metastable region, is an important observation. The long-lived metastable regime can be easily mistaken for the steady state in experiments or simulations. Furthermore, although not evident from the images presented, the Langevin simulations are becoming unstable for $\tau\sim10^5$ due to the accumulation of numerical errors. The simple model does not suffer from this problem and has the benefit of being directly interpretable.    

\section{Conclusion}

In conclusion, we have presented a theoretical study of the stochastic
dynamics of a particle in the periodic double-well potential. A combination of the matrix continued fraction technique applied to solve the Fokker-Planck equation combined with the full counting statistics and simple master equation models allowed us to determine the role of particular processes in the overall dynamics. 

For strong damping, analysis of the velocity Fano factor revealed a region where
single jumps, including fractional ones, are the prevailing
source of the velocity noise. We have shown that with decreasing temperature, the steady-state properties of the overdamped junction approach an effective single well system for any finite $\alpha$.

In the intermediate and weak damping regime, the FCS analysis showed complex dynamics related to the single and multiple positional jumps. The revealed large differences between the rates of odd (fractional) and even positional jumps can be explained by analyzing the particle trajectories. Even in this regime, we have identified parameter ranges, for which a single-harmonic analysis is sufficient for the description of the main positional slip statistics.  

In the analysis of the switching regime, we have presented a simple master
equation method to calculate the escape and retrapping rates. We have focused on both the regime where the double-well character plays an important role in the steady state statistics and the regime where it does not. We have shown how this property is related to the retrapping, escape, and positional jump rates.  We have demonstrated how these rates evolve with the bias force and determined the probabilities of steady-state occupations. The important observation is that retrapping, escape, and positional jumps can have rates that differ by orders of magnitude. This sets distinct time scales in realistic retrapping processes.

To illustrate this, we have investigated a quench, where the system in the fully running steady state is quenched to a bias force near the critical lower retrapping bias. We have shown that the results of the simplified model are in agreement with full Langevin simulations at longer times and that the differences at the short times are due to the details of the initial state used. The advantage of the simplified model is its stability at long times and, more importantly, its straightforward interpretability. Each time regime can be related to particular rates. In this way, we have been able to identify a metastable locked state, which can be easily mistaken for a steady state, due to the low escape rates and low rates of the jumps between the minima. 

To wrap up, besides providing a simple technique for analyzing the statistical properties of stochastic systems with double-well periodic potentials and analyzing their properties, we have shown two rather general features of such systems that can be crucial for the analysis of experimental setups. First, for a broad range of parameters in a strong and intermediate damping regime, the steady-state system can be approached with a simple single-harmonic model. Consequently, the two-well character of real potentials can be hidden in the averaged data when the escape rate from one of the minima significantly exceeds the other rates. This can be true even for wells of similar depths at low noise.

On the other hand, the occupations of the two minima in the fully locked steady state can be significantly different even for nearly equal minima if the damping is low. In addition, these occupations are highly parameter sensitive.  
What is also important for analysis of experiments and numerical simulations is the realization that the retrapping time and the time when the steady-state occupation is finally reached can differ by several orders of magnitude.

\emph{Acknowledgments.\textemdash{}}  M.\v{Z}. and T.N. acknowledge support by the Czech Science Foundation through Project No. 23-05263K. This work was supported by the Ministry of Education, Youth and Sports of the Czech Republic through the e-INFRA CZ (ID:90254). E.G. thanks Deutsche Forschungsgemeinschaft (DFG Project No. GO-1106/6) for financial support of this collaboration. W.B. was financially supported by the Deutsche Forschungsgemeinschaft (DFG, German Research Foundation) via the SFB 1432 (Project ID~425217212).  M.\v{Z}. thanks R\'obert Jur\v{c}\'ik for rewriting the code for stochastic dynamics into Julia.   

\input{paper.bbl}

\end{document}

%% file: paper.bbl
%